\documentclass[preprint]{aastex}
\include{amssymb}
\usepackage{amsmath}

\defcitealias{sawalaetal2010}{SSM10} 

\begin{document}
\title{{\bf The ACS LCID Project XI. On the early time resolution of LG dwarf galaxy SFHs: 
Comparing the effects of reionization in models with observations}\altaffilmark{*}}

\author{Antonio Aparicio\altaffilmark{1,2},
Sebastian L. Hidalgo\altaffilmark{1,2},
Evan Skillman\altaffilmark{3},
Santi Cassisi\altaffilmark{4,1},
Lucio Mayer\altaffilmark{5,6},
Julio Navarro\altaffilmark{7},
Andrew Cole\altaffilmark{8},
Carme Gallart\altaffilmark{1,2},
Matteo Monelli\altaffilmark{1,2},
Daniel Weisz\altaffilmark{9,10,11}
Edouard Bernard\altaffilmark{12},
Andrew Dolphin\altaffilmark{13} and
Peter Stetson\altaffilmark{14}
}

\altaffiltext{*}{Based on observations made with the NASA/ESA Hubble Space Telescope, 
obtained at the Space Telescope Science Institute, which is operated by the 
Association of Universities for Research in Astronomy, Inc., under NASA contract NAS 
5-26555. These observations are associated with program \#10505}
\altaffiltext{1}{Instituto de Astrof\'\i sica de Canarias. V\'\i a L\'actea s/n.
E38200 - La Laguna, Tenerife, Canary Islands, Spain;
aaj@iac.es, shidalgo@iac.es, carme@iac.es, monelli@iac.es}
\altaffiltext{2}{Department of Astrophysics, University of La Laguna. V\'\i a L\'actea s/n.
E38200 - La Laguna, Tenerife, Canary Islands, Spain}
\altaffiltext{3}{INAF-Osservatorio Astronomico di Collurania, Teramo, Italy; cassisi@oa-teramo.inaf.it}
\altaffiltext{4}{Intitut f\"ur Theoretische Physics, University of Zurich, Z\"urich, Switzerland; lucio@physics.unizh.ch}
\altaffiltext{5}{Department of Physics, Institut f\"ur Astronomie, ETH Z\"urich, Z\"urich, Switzerland; lucio@phys.ethz.ch}
\altaffiltext{6}{Department of Physics and Astronomy, University of Victoria, BC V8P 5C2, Canada; jfn@uvic.ca}
\altaffiltext{7}{Minnesota Institute for Astrophysics, University of Minnesota, Minneapolis, MN 55455, USA; skillman@astro.umn.edu}
\altaffiltext{8}{School of Physical Sciences, University of Tasmania, Hobart, Tasmania, Australia; andrew.cole@utas.edu.au}
\altaffiltext{9}{Astronomy Department, Box 351580, University of Washington, Seattle, WA 92895, USA; dweisz@uw.edu}
\altaffiltext{10}{Department of Astronomy, University of California at Santa Cruz, 1156 High Street, Santa Cruz, CA 95064, USA}
\altaffiltext{11}{Hubble Fellow}
\altaffiltext{12}{Institute for Astronomy, University of Edinbourgh, Royal Observatory, Blackford Hill, Edinbourgh EH9 3HJ, UK; ejb@roe.ac.uk}
\altaffiltext{13}{Raytheon, 1151 East Hermans Road, Tucson, AZ 85706, USA; adolphin@raytheon.com}
\altaffiltext{14}{Dominion Astrophysical Observatory, Herzberg Institute of Astrophysics, National Research Council, 5071 West Saanich Road, Victoria, British Columbia V9E 2E7, Canada; peter.stetson@nrc-cnrc.gc.ca}

\begin{abstract}

The analysis of the early star formation history (SFH) of nearby galaxies, obtained from 
their resolved stellar populations is relevant as a test for cosmological models. 
However, the early time resolution of observationally derived SFHs is limited by several 
factors. Thus, direct comparison of observationally derived SFHs with those derived from 
theoretical models of galaxy formation is potentially biased. Here we investigate and 
quantify this effect. For this purpose, we analyze the duration of the early star formation 
activity in a sample of four Local Group dwarf galaxies and test whether they are consistent 
with being true fossils of the pre-reionization era; i.e., if the quenching of their star
formation occurred before cosmic reionization by UV photons was completed.
Two classical dSph (Cetus and Tucana) and two dTrans (LGS-3 and Phoenix) isolated galaxies 
with total stellar masses between $1.3\times 10^6$ to $7.2\times 10^6$ M$_\odot$ have been 
studied.  Accounting for time resolution effects, the SFHs peak as much as 1.25 Gyr earlier 
than the optimal solutions.  Thus, this effect is important for a proper comparison of
model and observed SFHs. It is also shown that none of the analyzed galaxies can be 
considered a true-fossil of the pre-reionization era, although it is possible that the 
{\it outer regions} of Cetus and Tucana are consistent with quenching by reionization.

\end{abstract}


\section{INTRODUCTION}\label{secint}

Dwarf galaxies are at the focus of a major cosmological problem affecting the $\Lambda$CDM scenario: the number of dark matter subhalos around Milky Way-type galaxies predicted by $\Lambda$CDM simulations is much larger than the number of observed satellite dwarf galaxies \citep*{kauffmannetal1993,klypinetal1999,mooreetal1999}. Most proposals to overcome this problem stem from the idea that the smallest halos would have formed very few stars or failed to form stars at all, and that gas 
would have been  removed in an early epoch. In this way, the lowest mass sub-halos would be either completely dark, and thus undetectable, or extremely faint. Two main mechanisms are usually 
invoked as responsible of the smallest sub-halos failing to have an extended star formation history (SFH): heating from cosmic ultraviolet (UV) background radiation 
arising from the earliest star formation in the universe \citep*{bullocketal2000} and internal SN feedback \citep{maclow&ferrara1999} following the first star formation episodes in the host 
dwarf galaxy. The cosmic UV background raises the entropy of the 
intergalactic medium around the epoch of reionization,
preventing baryons from falling into the smallest sub-halos and it 
can also heat and evaporate the interstellar medium of larger sub-halos which have managed 
some star formation. 
The former would never form stars while the latter would presently show only a very old stellar population 
\citep{maclow&ferrara1999,sawalaetal2010,shenetal2014,benitezllambayetal2015}. Recent high resolution simulations of dwarf galaxy formation show that
the cosmic UV radiation field can also still suppress star formation, even when it cannot evaporate the gas from the halos, by simply preventing gas from becoming
dense enough to form molecular clouds \citep{shenetal2014}, verifying a previous proposal by \citet{schaye2001}.

It has also been proposed that ram pressure stripping in the
diffuse corona of the host massive galaxy could very rapidly remove the ISM already heated by the cosmic UV even over a large range of dwarf galaxy masses \citep{mayeretal2007}. However, such a
mechanism would become dominant later, during the main accretion phase of typical Milky Way-sized halo, at $z < 2$, in principle allowing star formation to extend for at least a couple of Gyr beyond 
the epoch of reionization.

Although consensus exists on the important role played by the two former mechanisms, 
less clear is the mass range of the affected sub-halos 
\citep*[e.g.,][]{gnedin2000,kravtsovetal2004,shenetal2014,benitezllambayetal2015}. 
The fact that most or all of the recently discovered ultra-faint dwarfs (UFDs) appear 
to host only a small population of very old stars, points to them as possible fossils of this 
process, but some of the classical dSph galaxies may also be affected. Besides 
heating the gas, UV photons produce the global cosmic reionization. 
The redshift at which the universe was fully reionized was $z\sim6$, as obtained from the 
presence of the Gunn-Peterson trough in quasars \citep{loeb&barkana2001,beckeretal2001},
although there is increasing evidence that this process was inhomogeneous 
\citep{spitler2012,becker2015,sobacchi2015}.  
According to models \citep[see e.g.,][]{ricotti&gnedin2005,gnedin&kravtsov2006,bovill&ricotti2011a,bovill&ricotti2011b}, the minimum circular velocity for a dwarf halo to accrete and cool gas in order to produce star formation is in 
the range of $v_c\sim20$ to $30$ km\,s$^{-1}$, which corresponds 
to a total mass of $\sim 10^8-10^9$ M$_\odot$. 
However, while most dwarf galaxies in the Local Group show circular velocities below 
this range and dynamical masses smaller than $\sim10^8$ M$_\odot$ 
\citep[see e.g.,][]{mcconnachie2012}, many of them have CMDs that have been interpreted 
as indicating the presence of star formation activity extended well beyond the 
reionization epoch, even in old dSph galaxies 
\citep{grebel&gallagher2004,monellietal2010a,monellietal2010b,hidalgoetal2011,weiszetal2014a,weiszetal2014b}. 
Two main mechanisms have been proposed to overcome this problem. 
The first one is that dwarf halos could have been much larger in the past and have lost a 
significant amount of mass due to tidal harassment \citep{kazantzidisetal2004,kravtsovetal2004}. 
This scenario is further supported by detailed simulations of the tidal interaction between satellites
and the host which includes also the baryonic component and ram pressure stripping \citep{mayeretal2007, mayer2010}. However, counter-arguments exists pointing to dwarf halos being resilient to tidal harassment \citep{penarrubiaetal2008}. The second one is that a self-shielding mechanism would be at work, protecting the gas in the central denser regions of the dwarf galaxy, where gas can be optically thick to the impinging radiation field \citep{susa&umemura2004}. The first mechanism is 
robust, since it is a natural consequence of 
hierarchical accretion as dwarf satellites move on highly eccentric orbits, suffering strong tidal shocks from the host potential. The second mechanism
is more subtle, since models neglect a related, competitive effect, namely that the local UV radiation from the primary galaxy or nearby proto-clusters
could have been much higher than the mean cosmic ionizing flux at $z > 1$ \citep{mayer2010,ilievetal2011}.

Models by \citet{bovill&ricotti2011a,bovill&ricotti2011b} show that Milky Way satellites with total luminosities $L_V>10^6 L_\odot$ are very unlikely to be true fossils of the reionization epoch, 
and that they probably are the result of hierarchical build-up from smaller halos extending beyond the reionization epoch. More specifically, \citet{bovill&ricotti2011a} conclude 
that the simulated properties of true fossils, i.e., those which have not undergone
any merging events after reionization, agree with those of a subset of the ultra-faint dwarf satellites of Andromeda and the Milky Way. Also, they found that most classical dSph satellites are 
unlikely true fossils, although they have properties in common with them: diffuse, 
old stellar populations and no gas.
We note that, while it would seem natural to associate UFDs to reionization fossils, alternative explanations for their origin have recently appeared in the literature in which at least a fraction of them could be remnants of the oldest, most heavily stripped population of galaxy satellites accreting at $z > 2$ onto the Milky Way halo \citep{tomozeiuetal2015}.

 \citet{bovill&ricotti2011b} used the fraction of star formation produced before reionization as a test to distinguish true fossil galaxies from non-fossils, defining the former as those having produced at least 70\% of their stars by $z=6$. This criterion was defined after the analysis of the theoretical simulations and has the advantage of allowing a thorough 
comparison of models and observations. In turn, \citet{weiszetal2014b} have obtained the SFHs of 38 Local Group dwarf galaxies with stellar masses in 
the range $10^4<M_\star <10^9 M_\odot$, finding that only five of them are consistent with forming the bulk of their stars before reionization and that only two out of the 13 predicted true fossils by \citet{bovill&ricotti2011b} show a star formation quenched by reionization. 
However, it should be noted that the results of \citet{weiszetal2014b} are affected by
limited time resolution at early ages while the  
predictions by \citet{bovill&ricotti2011b} are free of the these effects.
These effects are expected to significantly modify the SFHs obtained from observational data, 
as can be seen in the 
simulations shown by e.g., \citet{hidalgoetal2011} or \citet{hidalgoetal2013} among others. 
As a consequence of this, either observational effects should be simulated in theoretical models 
or they should be taken into consideration when comparing observational results 
with the theoretical models. The second is the objective of this paper. 

In this paper, we discuss how the temporal resolution limitations of a SFH derived from a 
CMD might be accounted for. 
In particular, the goal of this paper is to obtain an estimate of the maximum 
fraction of mass, {\it consistent with the observations}, which has been converted into stars 
by $z=6$ and of the time by which 70\% of the 
baryonic mass has been converted into stars in each galaxy. 
These values can be directly compared with the predictions of any theoretical model of star formation in dwarf galaxies and we do this comparison, for illustrative purposes, with some of the most recent ones, currently available in the literature. We have used high resolution SFHs obtained for a set of isolated Local Group dwarfs by the LCID collaboration \citep[][and references therein]{hidalgoetal2013}. Our approach opens a new window to the possibility of testing the effects that cosmic ultraviolet (UV) radiation and internal SN feedback have on the early SFH of dwarf galaxies
(whether or not combined with self-shielding or other effects). 
This is an exploratory paper in which we provide details of the method and apply it to a sample of very well studied dwarf galaxies. 
Environmental effects, such as tides and ram pressure, affect the mappings between 
present-day and pre-infall mass satellites and between baryon content and star formation. 
Therefore, we specifically concentrate on a sample of relatively isolated faint dwarfs. The chosen sample is diverse enough to contain both dIrrs, transition dwarfs and dSphs. In future works, the method will be applied to an extended sample including ultra faint dwarfs.

The structure of the paper is as follows. 
In Section \ref{secdat} observational data are described. 
In Section \ref{secdeconvol}, the proposed method is explained and applied to a sample of four Local Group dwarf galaxies. 
SFHs from Section \ref{secdeconvol} are compared in Section \ref{secmod} with a 
representative set of state-of-the-art theoretical models of galaxy formation. 
The results are discussed in Section \ref{secsum}, together with the main conclusions. 
As with the previous LCID papers, cosmological parameters of 
$H_0=70.5$ km\,s$^{-1}$\,Mpc$^{-1}$, $\Omega_m=0.274$, and a flat Universe with 
$\Omega_\Lambda=1-\Omega_m$ are assumed \citep{komatsuetal2009}.

\section{DATA SELECTION}\label{secdat}

For this work, regarding the observational material, we have used the SFHs of the galaxies 
Cetus \citep*{monellietal2010a}, Tucana \citep*{monellietal2010b}, 
LGS-3 \citep*{hidalgoetal2011}, and Phoenix \citep*{hidalgoetal2009}, obtained by the LCID 
collaboration. Data were obtained with the ACS and WFPC2 onboard HST. 
The SFHs used in this paper were derived using the IAC method, based on the suite of 
codes IAC-star/IAC-pop/MinnIAC \citep{aparicio&gallart2004,aparicio&hidalgo2009,hidalgoetal2011}. 
For a more detailed analysis, we have divided the three galaxies with larger field coverages, 
 Cetus, Tucana, and LGS-3, into two regions: an inner one, within one scale length from the center, and an outer one located beyond two scale lengths.

The properties of the galaxies are summarized in Table \ref{t1}. The total mass in stellar objects ($M_\star$), $V$ luminosity ($L_V$), velocity dispersion ($\sigma$), and metallicity ($[Fe/H]$) are given. 
$L_V$, $\sigma$, and $[Fe/H]$ are from \citet{mcconnachie2012}. 
The $M_\star$ values are calculated scaling $L_V$ with the mass-luminosity relation obtained from the SFH solution of each galaxy. 
The $L_V$ values range from $0.5\times 10^6$ to $2.6\times 10^6 L_\odot$, 
bracketing the limiting value obtained by \citet{bovill&ricotti2011a,bovill&ricotti2011b} 
for galaxies likely to be true fossils.

\section{IMPROVING TIME RESOLUTION OF THE SFH AT OLD AGES}\label{secdeconvol}

Robust SFHs are derived from the CMDs of resolved stellar populations, but they are still 
affected by several sources of uncertainty that limit time resolution. In short, these 
sources are of three kinds: (1) those affecting data, (2) those linked to physical properties, 
and (3) those inherent to the methodology used to derive the SFH. Limitations of the first kind are
related to photon counting statistics, defective flat-field corrections, or PSF sampling, 
among others. Sources of the second kind 
are mainly distance to the object (contributing to blending and crowding), 
background and foreground contamination, and differential reddening. 
The term {\it observational effects} has often been used by our team to refer to these two 
kinds combined \citep[see e.g.,][]{aparicio&hidalgo2009} and we will adopt it here from now on. 
They result in limiting the photometry depth and completeness, which in turn vary with 
stellar colors and magnitudes. They are modeled (e.g., with artificial star tests) and accounted for when synthetic populations are computed.
Effects of the third kind refer to the robustness of the method 
and include the accuracy of the stellar evolution libraries from which synthetic populations 
are simulated as well as the way in which the best solution is reached. Other items closely related to the physics of the problem like the age-metallicity degeneracy in the CMD are also involved. 
All these effects combined result in time resolution limitations and age inaccuracies in the 
final SFH solutions \citep{aparicioetal1997, aparicio&hidalgo2009, hidalgoetal2011}. 

To derive the SFH from the CMD of a resolved stellar 
population one makes a reliable simulation of observational effects (1 and 2 above) 
in the CMD of one or several synthetic stellar populations. 
These CMDs are in turn compared with the observational CMD to obtain the SFH 
\citep[see e.g.,][and references therein]{aparicio&hidalgo2009, hidalgoetal2011}. 
\citet{aparicio&hidalgo2009} and \citet{hidalgoetal2011} made an analysis of the effects of this process 
on the time resolution. To a good approximation it can be assumed that all together, the effects on the SFH are 
similar to a temporal shift plus a convolution with a Gaussian function, $G_{obs}$. {\bf It should be noticed that average ages of older populations tend to be biased to younger values. This is mainly the effect of the time limit imposed by the method, that does not allow ages older than the age of the universe in the solution. 

The function $G_{obs}$ can be obtained} according to the following recipe \citep[see][]{hidalgoetal2011}. 
First, a single burst stellar population with no age dispersion (or a very small one) is 
computed with the age at which $G_{obs}$ is searched. Second, the observational effects 
obtained for the galaxy in the artificial star tests are simulated. Third, the SFH of 
this synthetic population is derived. The result can be taken as 
$G_{obs}$. 

The $G_{obs}$ function can be used to partially remove the observational effects from the SFH derived for a galaxy. To do so, one can proceed parametrically. We do this by: (i) computing 
a large number of model SFHs of given shape, peak, and duration; (ii) convolving them 
with $G_{obs}$, and (iii) selecting those producing results compatible with the 
observationally derived SFH. The result of this inverse procedure is not the real, 
free of errors SFH. The problem we are facing is in fact a bias-variance tradeoff one \citep[see e.g.,][]{hastieetal2009}. Our objective is removing bias, but the intrinsic variance of the problem remains. Nevertheless it provides a better SFH for comparing to the theoretical models of galaxy formation.  This is especially true for the oldest ages, where the limitations on temporal resolution are greatest.

Specifically, we proceed as follows for each observed galaxy field. 
First, we obtain $G_{obs}$ functions for a range of input ages from 
5.0 to 13.0 $Gyr$ and we select the one whose resulting peak age is closest 
to the galaxy SFH peak age. Input ages of the selected $G_{obs}$ functions 
range from 11.0 (inner LGS-3 field), to 13.0 Gyr (outer Cetus and Tucana fields). 
Second, we have computed a large number of SFHs shaped according to Gaussian 
functions of amplitude $y$, mean age $\tau$, and standard deviation $\sigma$. 
We will refer to these trial model SFHs simply as {\it trial models}. 
Values of $\tau$ have been sampled 
from 6.0 to 13.5 Gyr with step of 0.1 Gyr. In turn, $\sigma$ has been sampled 
from 0.1 to 6 Gyr with steps of 0.1 Gyr and with the additional criterion that 
the resulting functions are truncated for age larger than 13.5 Gyr. Trial models include a simple but realistic simulation of the metallicity distribution of the real galaxy for the model age. To this, a gaussian metallicity distribution is used. The mean metallicity is the metallicity observationally obtained for the galaxy at age $\tau$ and the sigma is the metallicity dispersion obtained for the galaxy and the same age. Each trial model shape has been computed a total of 201 times, each one with a 
different value of $y$. A total of about $10^6$ trial models have been computed for 
each galaxy field. 

In the third step, each trial model has been convolved with the selected 
$G_{obs}$ function. We will refer by {\it convolved models} to the results of these 
convolutions. The fourth step has been to select all the convolved modles that are 
compatible with the SFH obtained for the galaxy field within the error intervals 
of the latter, including its integral. To apply this criterion, only the age 
interval $13.2 \leq {\rm age} \leq 10.0$~Gyr has been considered. The final step 
has been to average all the compatible convolved modles and, in turn, all the 
trial models originating them. 

\begin{figure}
\centering
\includegraphics[width=14cm,angle=0]{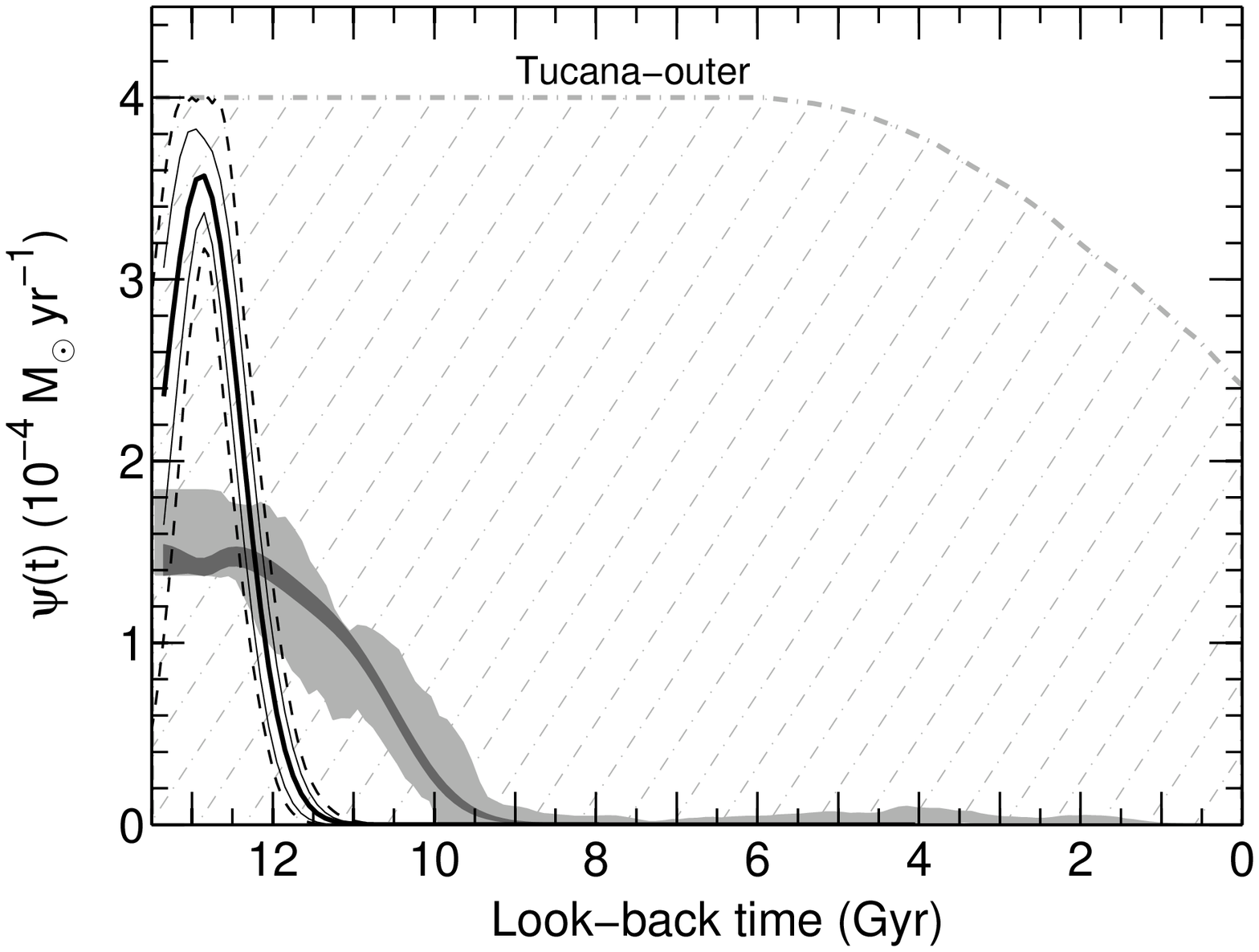}
\protect\caption[ ]{Summary of the used models and the good ones selected for the case of Tucana outer field. The grey dashed area shows the envelope of all (about $10^6$) the used models. The light grey region shows the SFH of the galaxy including the 1$\sigma$ error interval estimate. The dark grey region shows the average of convolved-SFHs compatible with empirical results. The wideness shows the 1-$\sigma$ dispersion. Thick, black line shows the average of the good trial models. Thin, black lines show the 1-$\sigma$ dispersion while thin dashed lines are the envelope of all the good trial models. 
}\label{f0}
\end{figure}

\begin{figure}
\centering
\includegraphics[width=14cm,angle=0]{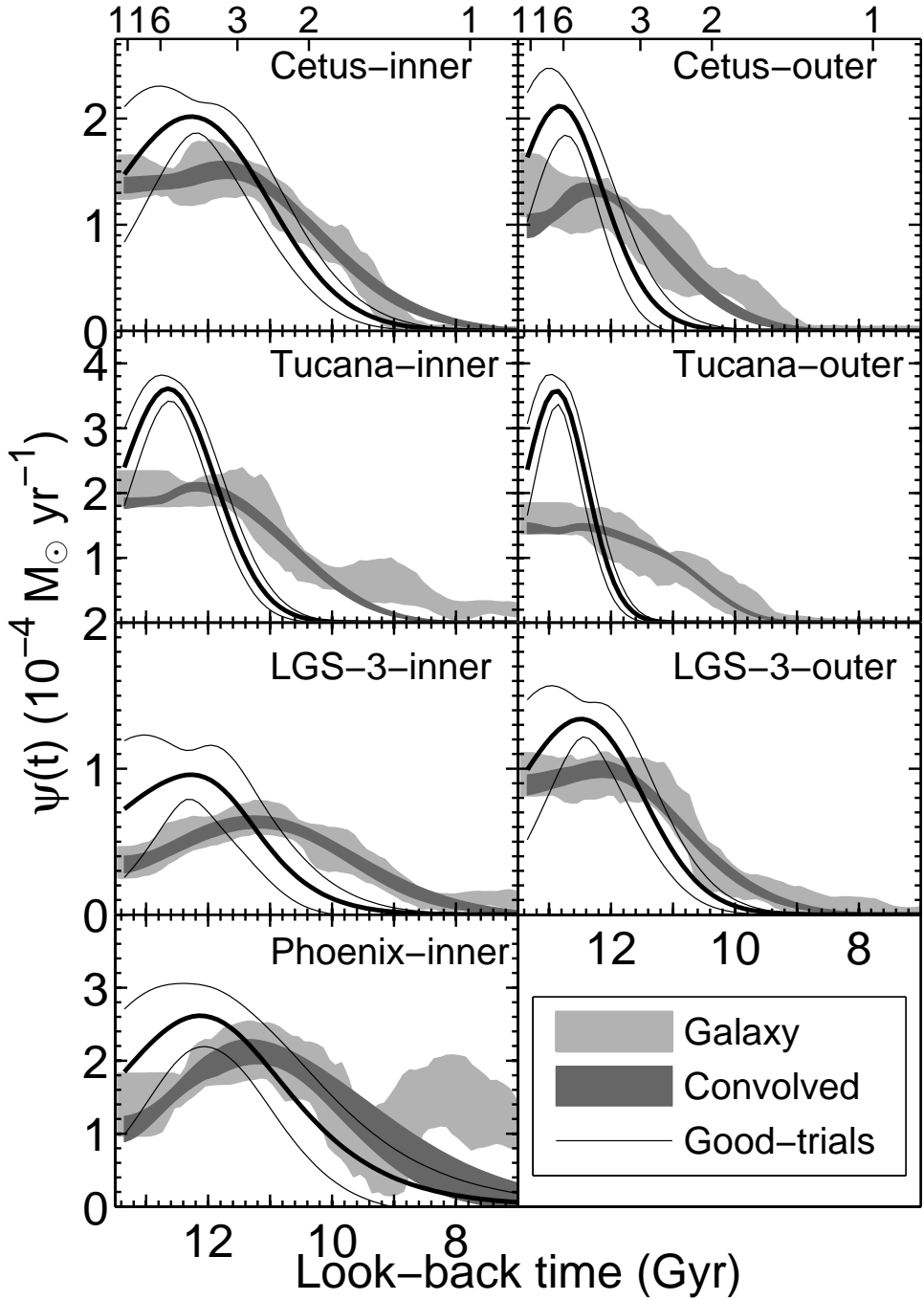}
\protect\caption[ ]{SFHs of the galaxy sample compared with good models. Inner and outer regions are shown for Cetus, Tucana and LGS-3 while only the inner region is considered for Phoenix. Light grey regions show the SFHs of galaxies, including the 1$\sigma$ error interval estimate. Dark grey regions show the average of convolved models compatible with empirical results. The wideness of these regions show the 1-$\sigma$ dispersion. Thick, black lines show the average of the good trial models. Thin, black lines show the 1-$\sigma$ dispersions. 
}\label{f1}
\end{figure}

Figure \ref{f0} shows a summary of the models, including the ones selected 
for the case of the outer field of the Tucana galaxy. General results are given 
in Figure \ref{f1}, which shows that, except for the case of Phoenix, 
the average of the {\it good} trial models peaks at older ages and is narrower than 
the convolved models or the observational SFHs. Following this, the averages of the good 
trial models will be used to make direct comparisons to theoretical models of galaxy formation.

The Gaussian functions used above are clearly a simplified representation of the 
SFH. It is not expected to reproduce extended SFHs or those in which 
recursive star formation bursts are going on for an extended period of time. Indeed, 
Figure \ref{f1} shows that intermediate ages for Phoenix are not well reproduced.
Nonetheless it can still be a good approximation for old stellar populations which 
is the purpose of this exercise.
To check for the dependence of the solutions with the trial model shape, 
we have repeated the same experiments using in turn triangular and step functions. 
In all the cases results are similar and do not significantly change the conclusions 
of our work. 

\begin{figure}
\centering
\includegraphics[width=14cm,angle=0]{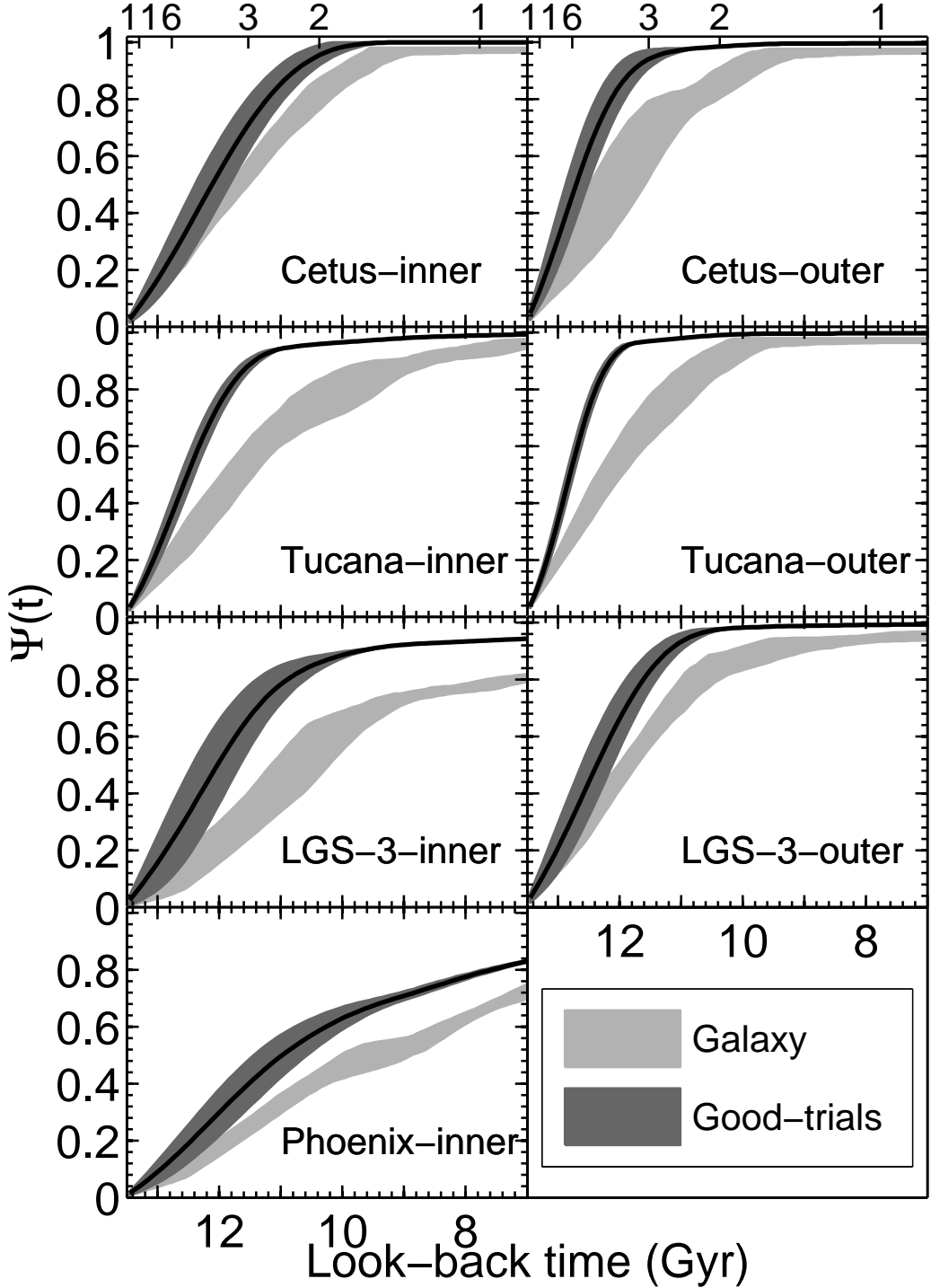}
\protect\caption[ ]{Cumulative SFHs for each galaxy field. Light grey areas show the cumulative SFHs derived from observations. Dark grey areas show the same for the average of the good trial models. See main text for details.
}\label{f2}
\end{figure}

Since we are interested here in the fraction of stars formed before a given time, 
it is better to use the cumulative SFH. We define it as $\Psi(t)=\int_T^t \psi(t')\,dt'$, 
where $T=13.5$ is the age assumed for the onset of the star formation and $\psi(t)$ is 
the star formation rate. Figure \ref{f2} shows the observed and the good 
trial model cumulative SFHs for all the galaxy fields. The latter is the information needed 
to make a direct comparison with theoretical models of galaxy formation.

\section{COMPARISON WITH THEORETICAL MODELS OF GALAXY FORMATION PREDICTIONS}\label{secmod}

\begin{figure}
\centering
\includegraphics[width=14cm,angle=0]{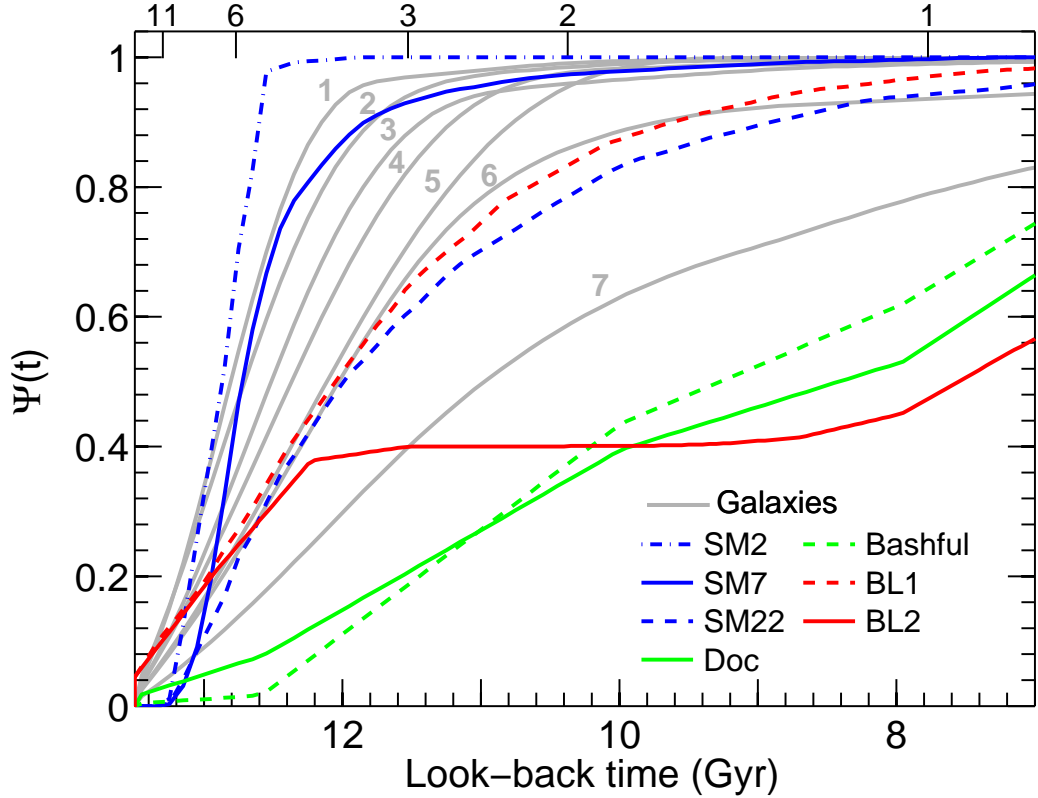}
\protect\caption[ ]{Averages of the good trial models in each galaxy field (plotted by thick, black lines in Figure \ref{f2}) compared with the same for a number of theoretical models of galaxy formation. Numbers in the figure correspond to 1: Tucana-outer, 2: Cetus-outer, 3: Tucana-inner, 4: LGS-3-outer, 5: Cetus-inner, 6: LGS-3-outer, and 7: Phoenix-inner. Regarding to models, SM2, SM7 and SM22 are respectively models 2, 7 and 22 by \citet{sawalaetal2010}. Model "DOC" is by \citet{shenetal2014}. Models "BL1" and "BL2" are models 1 and 2 by \citet{benitezllambayetal2015}.
}\label{f3}
\end{figure}

In Figure \ref{f3} the average of the good trial models of the galaxies are compared with a few, representative theoretical models of galaxy formation, namely models 2, 7, and 22 by \citet{sawalaetal2010}, 
model {\it Doc} by \citet{shenetal2014}, and two models by \citet{benitezllambayetal2015} representative of old and old+intermediate age dwarf galaxies. 
The latter are those given by the authors by the green curves in their figure 5. 

\citet{sawalaetal2010} have presented high-resolution hydrodynamical simulations of the formation and evolution of isolated dwarf galaxies including the most relevant physical effects, namely, metal-dependent cooling, star formation, feedback from Type II and Ia SNe, UV background radiation, and internal self-shielding. Models 2 and 7 include UV background radiation and internal self-shielding together with SNe feedback. Model 22 includes only SNe feedback, i.e., it is representative of the case that UV background radiation has no effect on the SFH. 

\citet{shenetal2014} have carried out fully cosmological, very high resolution, $\Lambda$CDM simulations of a set of field dwarf galaxies. Model {\it Doc} corresponds to a virial mass of $M_{vir}=1.16\times 10^{10} M_\odot$. Finally, \citet{benitezllambayetal2015} have used the cosmological hydrodynamical simulation of the Local Group carried on as part of the CLUES project. In all the cases, the reader is referred to the source papers for details.

Some relevant properties of the model galaxies are summarized in Table \ref{t1}. The total mass in stellar objects ($M_\star$), $V$ luminosity ($L_V$), velocity dispersion ($\sigma$), and metallicity ($[Fe/H]$) are given. $M_\star$ is provided by \citet{sawalaetal2010} and \citet{shenetal2014}, respectively. Regarding $L_V$, the values given by \citet{shenetal2014} are listed for their models. 
For the \citet{sawalaetal2010} models, the $L_V$ are obtained with IAC-star using the model SFHs provided by these authors as input. 
No data are given for the \citet{benitezllambayetal2015} models, since these authors do not provide them. 

Figure \ref{f3} shows that the average of the cumulative good trial models of Cetus, Tucana, and LGS-3 lay between models SM2 and SM7 (both including reionization), on the one side, and models SM22 (without reionization) and BL-1 (average of their oldest models), on the other side.  
Furthermore, the outer parts of Cetus and Tucana show a reasonable correspondence with SM7, indicating that reionization could have played a role in quenching the star formation in those regions. 
However, according to this figure, the inner regions of both galaxies plus LGS-3 (inner and outer) and Phoenix seem difficult to reconcile with the \citet{sawalaetal2010} models with reionization. A good correspondence exists between the inner region of LGS-3 and models BL-1 and SM22, while no good match is found for Phoenix with any of the models considered here. 
Finally, models BL2, Doc, and Bashful are, by far, too young to reproduce any of the observed galaxies and are likely to be more suitable for much younger galaxies, like IC-1613 \citep[see][]{skillmanetal2014}.

\citet{bovill&ricotti2011b} have defined true fossil galaxies as those having formed all or most of their stars before the reionization era, at $z=6$. They used the cumulative fraction of star formation produced before reionization as a test to distinguish true fossil galaxies from non-fossils, defining the former as those having produced at least 70\% of their stars at $z=6$, which, for the cosmological parameters adopted here (see above), corresponds to  approximately $12.8$ Gyr ago. Table \ref{t2} gives, for each galaxy field and model, the age at which $\Psi(t)=0.7$ and the value of $\Psi(t)$ for redshift $z=6$. Columns 2 and 3 list the values corresponding to the observed SFHs. Columns 4 and 5 give those corresponding to the average of the good trial models. Errors have been obtained from the error bands shown in Figure \ref{f2}. 

Two main conclusions can be obtained from Table \ref{t2}. First, the age for which the cumulative SFH, $\Psi(t)$ reaches 70\% is increased by observational effects by $\sim1.25\; Gyr$ on average, while the value of $\Psi(t)$ at $z=6$ is increased by a factor of $\sim 1.65$, on average, although it can be larger than two in some cases. This shows that working with the average of good trial models is useful if the earliest evolution of dwarf galaxies is sought. 
Second, taken at face value, none of the four galaxies analyzed here fulfill the criterion by \citet{bovill&ricotti2011b} of having $\Psi(t)\geq 0.7$ at $z=6$, to be considered true fossils from the pre-reionization era, although the outer regions of Cetus and Tucana might marginally 
qualify, within a 2$\sigma$ error interval. 
In addition, \citet{bovill&ricotti2011b} also concluded that galaxies with $L_V>10^6\,L_\odot$ are unlikely true fossils, while those with $L_V<10^6\, L_\odot$ remain reasonable candidates to be fossils of the first galaxy generation. 
However, three out of the four galaxies considered here have $L_V\leq 10^6 L_\odot$ 
(Tucana, LGS-3, and Phoenix) and are not true-fossils.

\section{SUMMARY AND CONCLUSIONS}\label{secsum}

In summary, we have presented a method to address the limitations of the temporal resolution of the early SFHs of galaxies derived from deep CMD modeling. We have applied the method to the analysis of the duration of the early star formation activity in a sample of Local Group dwarf galaxies with the purpose of testing whether or not they are true fossils of the pre-reionization era. For a study of this kind to be accurate, the affects of limited time resolution need to be accounted for. 
For testing for fossils, using SFHs directly derived from deep CMD modeling may produce 
biased results due to limited time resolution.
We have shown elsewhere \citep{hidalgoetal2011} that the SFH derived for a synthetic, 
single age and metallicity stellar population can be simulated by a Gaussian whose sigma 
depends on age ($G_{obs}$). 
The method presented here consists in computing a large number of stellar populations with 
Gaussian SFHs and metallicity distribution similar to the on obtained for the galaxy for the same age interval (that we call {\it trial models}; about $10^6$ per galaxy in this case) of varying mean, 
sigma, and amplitude, and selecting all those that, convolved with $G_{obs}$, produce results 
compatible with the optimal SFH directly derived from observations. 
The average of the good trial models is an improved approach to the real SFH of the galaxy
at the very earliest times. 
In general, the ages of the averages of the good trial models are about $\sim 1.25$\ Gyr older 
than the optimal SFH solutions for the case of predominantly old galaxies.

We have applied our method to four Local Group dwarfs: Cetus, Tucana, LGS-3, and Phoenix and 
we have compared our results with predictions made by a number of recent cosmological 
hydrodynamical simulations for the formation and evolution of dwarf galaxies. 
A relatively sharp exhaustion of the star formation is necessary at an early epoch, 
close to $z=6$, to account for the SFH obtained for the outer region of Tucana and Cetus. 
The inner part of both galaxies, as well as LGS-3 and, more clearly, Phoenix, 
are compatible with models predicting more extended star formation activity for dwarf galaxies. 
However, none of the galaxies and fields studied here, except perhaps the outer regions 
of Cetus and Tucana, fulfill the criterion to be considered as fossil remnants of the 
pre-reionization era. As a cautionary note, at present, it cannot be excluded that Cetus and Tucana were affected by environmental effects to some extent. 
Their distances to the massive spirals of the LG, while placing them outside the virial radius 
of either today, is still small enough to admit scenarios in which these galaxies are on nearly 
radial orbits and they may have already had a close pericenter passage with their host 
\citet{kazantzidisetal2011}. 
If they are bound satellites, their relatively large velocity dispersion, 
$\sigma \sim 15$ km/s (see Table \t1), would make them comparable to the brightest dSph 
satellites such as Fornax, and implies their halos could have been much more massive 
before tides began to prune them substantially. If this were the case, it would be likely 
that their halo mass before infall was high enough to place them clearly above the threshold 
mass for reionization to play a role, strengthening further the conclusion that they cannot 
be reionization fossils.

\begin{acknowledgements}

The authors thanks the anonymous referee for her/his comments and suggestions, that help to improve the paper clarity.
This work has been funded by the Economy and Competitiveness Ministry of the Kingdom of Spain (grants AYA2010-16717 and AYA2013-42781-P) and by the Instituto de Astrof\'\i sica de Canarias (grant P3/94).
\end{acknowledgements}

\clearpage

\clearpage

\clearpage

\clearpage
%


\begin{deluxetable}{ccccc}
\tabletypesize{\scriptsize}
\tablecaption{Summary of data and models\label{t1}}
\tablewidth{0pt}
\tablehead{
\colhead{Galaxy/Model} &\colhead{$M_\ast$} &\colhead{$L_V$} &\colhead{$\sigma$}      &\colhead{$[Fe/H]$}\\
\colhead{} &\colhead{($10^6 M_\sun$)} &\colhead{($10^6 L_\sun$)} &\colhead{($km s^{-1}$)}  &\colhead{}\\
\colhead{(1)} & \colhead{(2)} & \colhead{(3)} & \colhead{(4)} & \colhead{(5)}}
\startdata
Cetus          & 7.17       &  2.58     &~~~~~~$17.0\pm 2.0$                         & $\,\,\,-1.9\,\,\,\pm 0.1 $\\                                               
Tucana         & 1.66       &  0.54     &~~~~$15.8\substack{+4.1 \\ -3.1}$           & $\,\,\,\,\,\,-1.95\pm 0.15$\\
LGS-3          & 2.08       &  0.94     &~~~~~\,$7.9\substack{+3.2 \\ -2.4}$         & $\,\,\,\,\,\,-2.10\pm 0.22$\\
Phoenix        & 1.29       &  0.78     &\nodata                                     & $\,\,\,-1.37\pm 0.2$\\
S2             & 0.96       &  0.43     &7.3                                         &$-1.76$~~~~~~~\\
S7             &10.02~\,    &  4.50     &9.1                                         &$-1.17$~~~~~~~\\
S22            & 2.58       &  1.34     &7.1                                         &$-1.03$~~~~~~~\\
Shen13-Doc     &34.0~~~     & 34.04~    &\nodata                                     &\nodata\\
Shen13-Bashful &115.0~~~~\, &135.52~~\, &\nodata                                     &\nodata\\
BL-1           &\nodata     &\nodata    &\nodata                                     &\nodata\\
BL-2           &\nodata     &\nodata    &\nodata                                     &\nodata\\
\enddata
\tablecomments{Column 1: data and models identification. Column 2: total stellar mass. Column 3: total $V$ luminosity. Column 4: central velocity dispersion. Column 5: metallicity. For galaxies, $L_V$, $\sigma$ and $[Fe/H]$ are from McConnachie 2012; The $M_\star$ values are calculated scaling $L_V$ with the mass-luminosity relation obtained from the SFH solution of each galaxy. $M_\star$ is provided by \citet{sawalaetal2010} and \citet{shenetal2014}, respectively. Regarding $L_V$, the values given by \citet{benitezllambayetal2015} are listed for their models. For \citet{sawalaetal2010} models, the $L_V$ are obtained with IAC-star using the model SFHs provided by these authors as input. No data are given for \citet{benitezllambayetal2015} models, since these authors do not provide them.}
\end{deluxetable}

\begin{deluxetable}{cccccccc}
\tabletypesize{\scriptsize}
\tablecaption{Age for $\Psi(t)=0.7$ and $\psi(t)$ value for $z=6$ \label{t2}}
\tablewidth{0pt}
\tablehead{
\colhead{Galaxy field} & \colhead{Age for $\Psi(t)$=0.7} & \colhead{$\Psi(t)$ at $z=6$} & \colhead{Age for $\Psi(t)$=0.7} & \colhead{$\Psi(t)$ at $z=6$}  \\
\colhead{} & \colhead{Observational} & \colhead{Observational} & \colhead{Good trials} & \colhead{Good trials} }
\startdata
Cetus inner     &$10.9\substack{+0.2 \\ -0.2}$      & $0.22\substack{+0.03 \\ -0.07}$    & $11.6\substack{+0.3 \\ -0.4}$   & $0.25\substack{+0.18 \\ -0.19}$ \\
Cetus outer     &$11.3\substack{+0.3 \\ -0.6}$      & $0.28\substack{+0.09 \\ -0.13}$    & $12.4\substack{+0.2 \\ -0.3}$   & $0.46\substack{+0.24 \\ -0.12}$ \\
Tucana inner    &$10.8\substack{+0.7 \\ -0.3}$      & $0.21\substack{+0.05 \\ -0.06}$    & $12.1\substack{+0.1 \\ -0.1}$   & $0.36\substack{+0.12 \\ -0.06}$ \\
Tucana outer    &$11.5\substack{+0.4 \\ -0.3}$      & $0.29\substack{+0.03 \\ -0.06}$    & $12.6\substack{+0.0 \\ -0.1}$   & $0.53\substack{+0.11 \\ -0.05}$ \\
LGS-3 inner     &$~\,9.4\substack{+0.3 \\ -0.5}$    & $0.10\substack{+0.04 \\ -0.06}$    & $11.4\substack{+0.4 \\ -0.4}$   & $0.23\substack{+0.25 \\ -0.13}$ \\
LGS-3 outer     &$11.2\substack{+0.3 \\ -0.2}$      & $0.23\substack{+0.05 \\ -0.06}$    & $11.9\substack{+0.3 \\ -0.2}$   & $0.31\substack{+0.22 \\ -0.11}$ \\
Phoenix inner   &$~\,7.3\substack{+0.3 \\ -0.2}$    & $0.08\substack{-0.03 \\ -0.03}$    & $9.1 \substack{+0.3 \\ -0.2}$   & $0.13\substack{+0.05 \\ -0.11}$ \\
SM2             &\nodata                            &\nodata            & $12.8$           & $0.98$ \\
SM7             &\nodata                            &\nodata            & $12.5$           & $0.67$ \\
SM22            &\nodata                            &\nodata            & $11.0$           & $0.31$ \\
Shen13-Doc      &\nodata                            &\nodata            & $~\,6.7$         & $0.08$ \\
Shen13-Bashful  &\nodata                            &\nodata            & $~\,7.3$         & $0.02$ \\
BL-old          &\nodata                            &\nodata            & $11.2$           & $0.34$ \\
BL-young        &\nodata                            &\nodata            & $5.8$            & $0.30$ \\
\enddata
\tablecomments{Column 1: Data source: observational inner or outer field for each galaxy or model; observational source: \citet{hidalgoetal2013}; models sources: \citet{sawalaetal2010}, \citet{shenetal2014} and \citet{benitezllambayetal2015}. Column 2: Age at which the cumulative SFH $\Psi(t)$ reaches 70\%. Column 3: Value of $\Psi(t)$ corresponding to $z=6$. Values listed in columns 2 and 3 have been obtained from Figure. \ref{f2}.}
\end{deluxetable}

\end{document}